# Bulk photonic metamaterial with hyperbolic dispersion

M. A. Noginov[1*], Yu. A. Barnakov[1], G. Zhu[1], T. Tumkur[2,3], H. Li[1], E. E. Narimanov[4]

[1] *Center for Materials Research, Norfolk State University, Norfolk, VA 23504*
[2] *Summer Research Program, Center for Materials Research, Norfolk State University, Norfolk, VA 23504*
[3] *Materials Engineering, Purdue University, West Lafayette, IN 47907*
[4] *Electrical Engineering, Purdue University, West Lafayette, IN 47907*

Photonic metamaterials are engineered nano-composite structures with optimized responses to electromagnetic fields at optical frequencies that lead to super[1]- and hyper[2-5]-lens and invisibility cloaks[6-12], bringing closer a variety of exciting applications and calling for inexpensive three-dimensional media. Most metamaterials are fabricated using prohibitively expensive nanolithogtaphy techniques and are available in a form of sub-wavelength thin films. Recent advances in non-lithographic fabrication include direct laser writing[13], epitaxy[14], self-organized[15] growth, and electroplating[16-18]. Here we demonstrate a self-standing 51µm-thick three-dimensional metamaterial based on the network of silver nanowires in an alumina membrane. This constitutes the anisotropic effective medium with hyperbolic dispersion[19], which can be used in sub-diffraction imaging[20] or optical cloaks[7]. Highly anisotropic dielectric constants of the material range from positive to negative, and the transmitted laser beam shifts both *toward* the normal to the surface, as in regular dielectrics, and *off* the normal, as in dielectrics with the refraction index smaller than one.

Commercially acquired self-supporting anodic alumina membranes (Fig. 1a) with the dimensions equal to 1cm x 1cm x 51µm, hole diameter equal to 35 nm and the porosity equal to 15% were filled with silver *via* the electroplating technique as described in Methods. Membranes were almost completely filled with silver at the electrode side, and the concentration of silver somewhat reduced toward the opposite side, Fig. 1b. In the best samples, the values of ε measured on both sides (determined from the reflectance studies described below) were close to each other, suggesting high uniformity of filling.

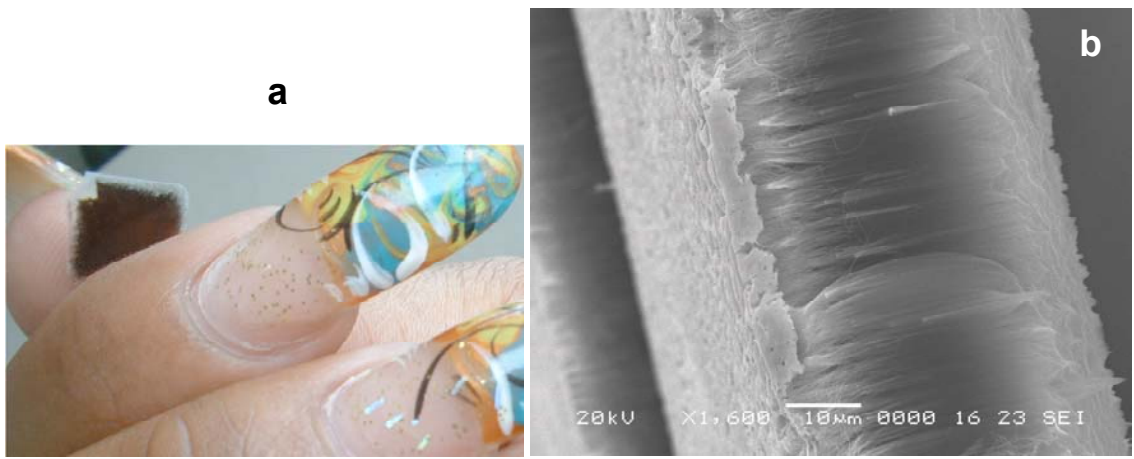

Figure 1. (a) Photograph of the silver-filled membrane. (b) Scanning Electron Microscope (SEM) picture of the etched side wall of the membrane showing loose silver nanowires.

[*] Corresponding author: *e-mail: mnoginov@nsu.edu; Phone: (757) 823 2204*

The sample's reflectance in *s* and *p* polarizations was measured as the function of the incidence angle at ten different wavelengths ranging from 458 nm to 950 nm, Fig. 2a. The spectrum of the experimentally measured Brewster angles (angles of minimal reflectance in *p*-polarization), Fig. 2b, shows a discontinuity at the wavelength ($\lambda=0.84$ µm) at which the dielectric constant in the direction perpendicular to the membrane's surface changes its sign from positive to negative, as shown in Fig. 3 and explained below. This pattern is qualitatively similar to that predicted theoretically and observed experimentally[14] in the multilayered sandwich structure of doped $In_{0.53}Ga_{0.47}As$ and undoped $Al_{0.48}In_{0.52}As$ semiconductor layers, which had hyperbolic dispersion law and demonstrated negative refraction of *p*-polarized light.

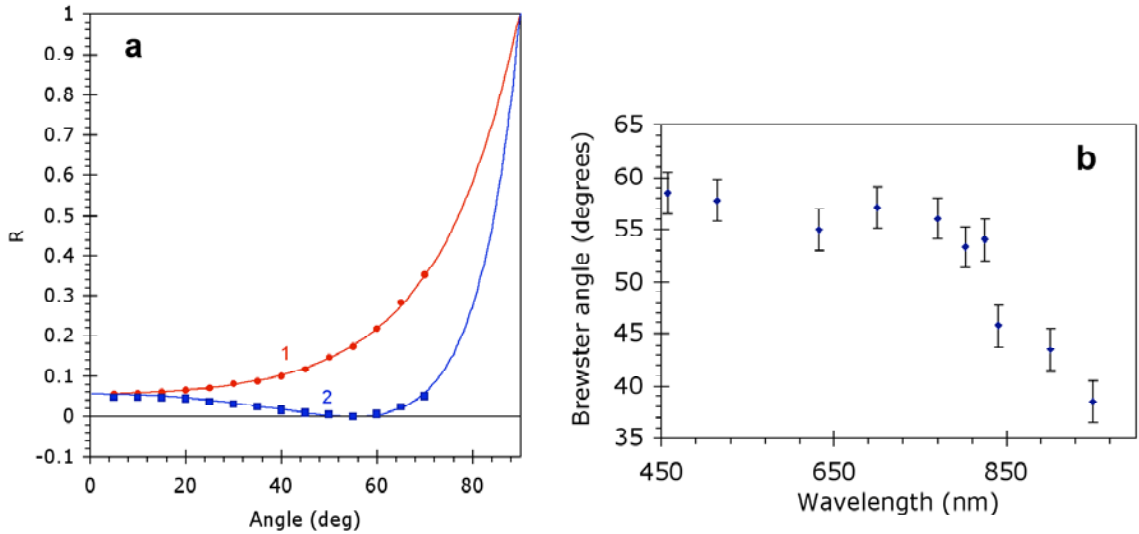

Figure 2. (a) The dependence of the reflectance on the incidence angle in *s*-polarization (trace 1, red circles) and *p*-polarization (trace 2, blue squares) at $\lambda=632.8$ nm. Solid lines – calculation according to Eq. 1. The scattering factor is taken into account (see Methods). (b) The summary of the experimentally measured Brewster angles.

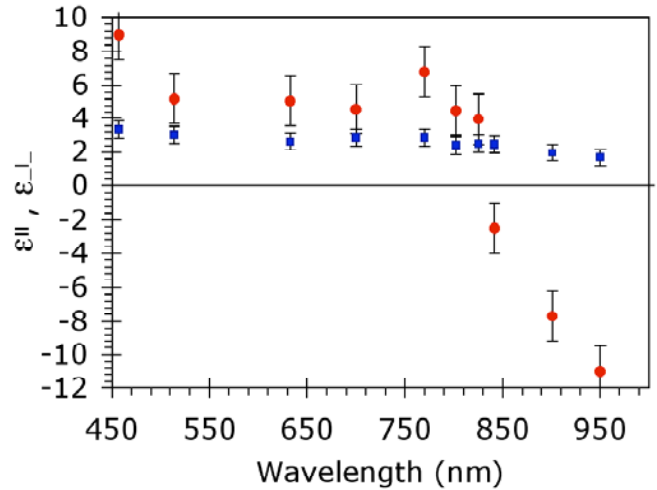

Figure 3. Real parts of $\varepsilon_\parallel$ (blue squares) and $\varepsilon_\perp$ (red circles) determined from the reflectance measurements.

The experimental curves were fitted with the known theoretical expressions for reflection from a uniaxial material

$$R=|r|^2 = \begin{cases} \left|\dfrac{\sin(\theta-\theta_t)}{\sin(\theta+\theta_t)}\right|^2, & \theta_t = \arcsin\left(\dfrac{\sin\theta}{\sqrt{\varepsilon_\parallel}}\right), & s \text{ polarization.} \\ \left|\dfrac{\varepsilon_\parallel \tan\theta_t - \tan\theta}{\varepsilon_\parallel \tan\theta_t + \tan\theta}\right|^2, & \theta_t = \arctan\sqrt{\dfrac{\varepsilon_\perp \sin^2\theta}{\varepsilon_\parallel\varepsilon_\perp - \varepsilon_\parallel \sin^2\theta}}, & p \text{ polarization.} \end{cases} \quad (1)$$

The real parts of the dielectric constants in the directions parallel and perpendicular to the membrane's surface, $\varepsilon_\parallel$ and $\varepsilon_\perp$, derived from the reflectance curves of Fig. 2a are shown in Fig. 3.

As expected, the synthesized metamaterial is highly anisotropic, with $\varepsilon_\perp$ changing sign from positive to negative at the wavelength of the discontinuity of the Brewster angle, Fig. 2b. According to Ref.[19], in silver-filled alumina membrane, $\varepsilon_\parallel$ is predicted to be positive and change slowly in the wavelength range studied, while $\varepsilon_\perp$ is expected to change sign at ~0.84 μm at the value of the filling factor equal to ~8%. The discrepancy of the filling factors, 15% vs 8%, can be explained by imperfect electroplating (see Methods) or the difference in properties of bulk silver[21] and the silver filled into the membrane.

Angles of refraction for both *p* and *s* polarized light were directly studied in silver-filled membranes at λ=632.8 nm in the modified knife edge setup of Ref.[14], Fig. 4a. First, the sample was positioned normally to the laser beam and the dependence of the transmitted intensity on the knife edge position was measured, Fig. 4b. Then, the knife edge was moved to the position where it blocked ~50% of the transmitted light. After that, the sample was rotated and, depending on the angle of refraction and the direction of rotation, the amount of light measured after the knife edge increased or decreased by a certain value, Figs. 4a,b.

We have shown that in *s* polarization, the directions of the beam shifts, as indicated by changes in the measured light intensity, were the same as in thin glass slide with the index of refraction *n* greater than one, Fig. 4c. At the same time, at *p* polarization, the beam shifted in the opposite direction, Fig. 4d. In isotropic media, such situation would correspond to *n*<1. This is another demonstration of the strong anisotropy of the synthesized metamaterial.

Dielectric constants greater than zero and smaller than one are needed in optical cloaks. This makes the developed silver-filled membranes promising for cloaking applications, in particular if the cylindrical geometry ("hair brush" structures[7]) can be achieved. The same geometry with the hyperbolic law of dispersion would be ideal for the realization of a hyperlens[20].

To summarize, we have demonstrated bulk photonic metamaterial based on anodized alumina membranes filled with silver. The material is highly anisotropic and follows the hyperbolic dispersion law at λ>0.84 μm. The refraction of light in the direction expected of isotropic media with n<1 has been experimentally demonstrated at λ=632.8 nm. The designed photonic metamaterial is the thickest reported in the literature, both in terms of its physical size, $d = 51$ μm, and the number of vacuum wavelengths, *N*=61 at λ=0.84 μm and *N*=81 at λ=0.63 μm. This combination of unique properties makes the material suitable for a variety of applications ranging from sub-diffraction imaging to optical cloaking.

When this work was completed, the paper has been published[22] reporting the negative refraction in thinner, $d \leq 11$ μm, membranes with comparable parameters.

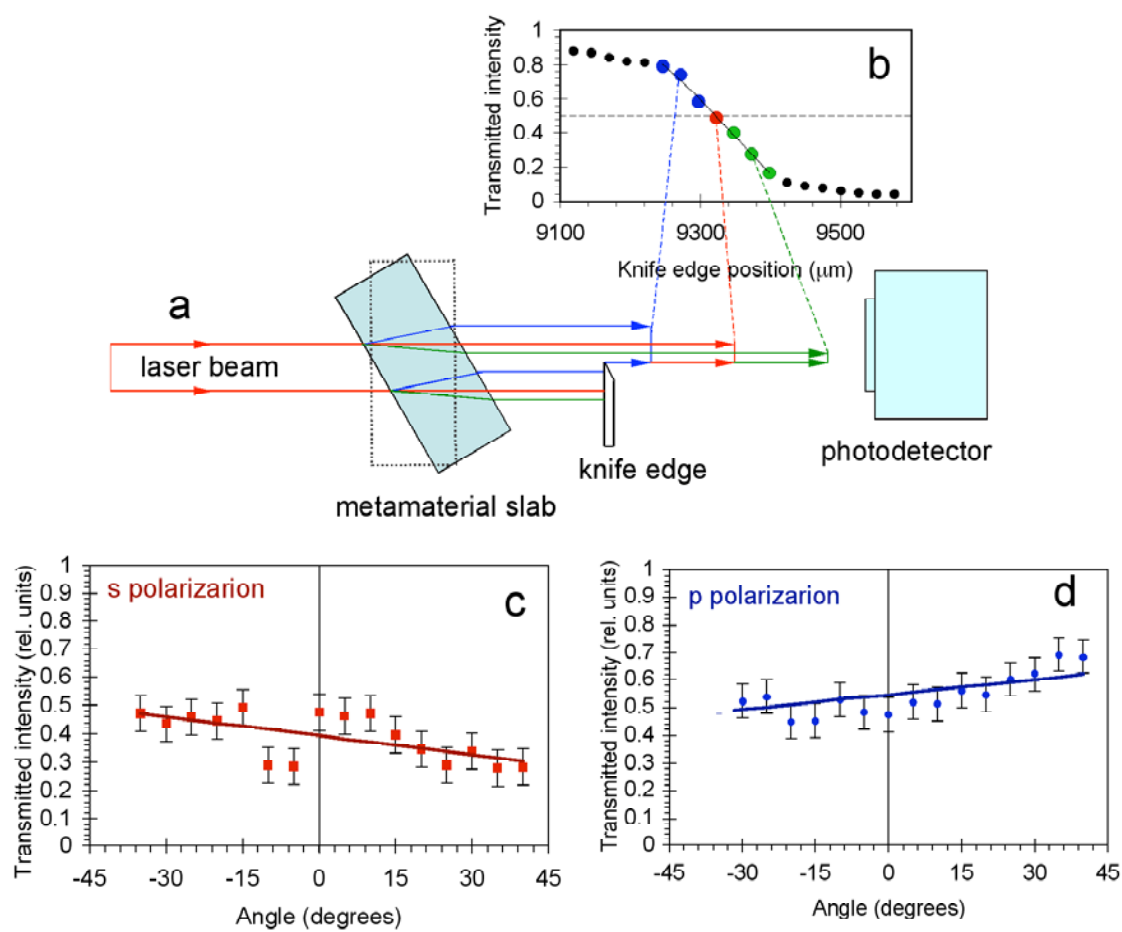

Figure 4. (a) Experimental setup used at the beam shift measurements; (b) beam profile measured with the knife edge technique; (c,d) shifts in opposite directions for *s* and *p* polarizations.

**Methods**

Synthesis

Anodic alumina membranes (AAMs) with the dimensions 1cm x 1cm x 51μm have been purchased from Synkera Technologies Inc. Silver nanowires were synthesized in AAMs *via* electrochemistry plating. Thin gold film (~50 nm) deposited on the membrane's surface *via* thermal vapor deposition technique served as the working electrode and the graphite rod played a role of the counter electrode. The mixture of aqueous solutions of Silver nitrate (1.76 M) and Boric acid (0.7 M) buffered with Nitric acid to pH 2-3 were used as electrolyte. The DC voltage of 1 V was applied during ~300 min deposition. Gold electrode was removed before the optical studies. (Note that the synthesized nanowires were two orders of magnitude longer than the subwavelengh nanorods reported in Refs.[16-18].)

Optical measurements

In studies of angular dependencies, the samples were mounted at the axis of rotation of the goniometer, where they were intercepted by the laser beam. The diameter of the laser beam was equal to ~0.5 mm in the reflectance measurements and ~0.15 mm in the beam shift measurements. Light transmitted by silver-filled membranes was strongly scattered. This contributed to the relatively large scatter of transmitted intensities, Fig. 4b, which was stronger in filled membranes than in pure membranes or in the thin glass slide. The transmitted intensities in Fig. 4b measured after the knife edge were normalized to those measured at the same angles without the knife edge.

Determination of $\varepsilon_\parallel$ and $\varepsilon_\perp$ from the reflectance curves

We first determined the values $\varepsilon_\parallel$ by fitting the reflectance in *s* polarization. Relatively small but noticeable amount of light was scattered diffusely. Consequently, the agreement between the fitting and the experiment strongly improved if the experimental data (in both polarizations) were scaled by the constant of the order of unity. Keeping $\varepsilon_\parallel$ and the scattering factor unchanged, we then used $\varepsilon_\perp$ as a variable parameter to fit the reflectance in *p* polarization. In the fitting procedure, we were mainly concerned with the position of the Brewster angle, since this parameter was experimentally known with the highest accuracy. In the two-step fitting procedure, the accuracy of $\varepsilon_\parallel$ was higher than the accuracy of $\varepsilon_\perp$. The reflectance curves were much more sensitive to real than to imaginary parts of $\varepsilon$. That is why only real parts of the dielectric constants, shown in Fig. 3, have been determined. According to the transmission measurements carried out in an integrated sphere setup at normal angle of incidence, the figure of merit (the ratio of real to imaginary parts of the index of refraction) for the direction parallel to the membrane surface exceeds 530 in the whole spectral range studied. This is consistent with the small effect of the imaginary part of $\varepsilon$ on the shape of the reflectance curves.

The work was supported by the NSF PREM grant # DMR 0611430, NSF CREST grant # HRD 0317722, NSF NCN grant # EEC-0228390, NASA URC grant # NCC3-1035 and ARO-MURI award 50342-PH-MUR. The authors thank Viktor A. Podolskiy for useful discussions and Natalia Noginova for the help with the data processing. Carla S. McKinney holds the sample in Fig. 1a.